\documentclass[prd,aps,preprintnumbers, showpacs, nofootinbib,superscriptaddress,notitlepage]{revtex4-1}
\usepackage{amsmath,amssymb,amsthm,amsfonts,amsbsy,latexsym}
\usepackage{array}
\usepackage{verbatim}
\usepackage{epsfig}
\usepackage{graphicx}
\usepackage{bm}
\usepackage{relsize}
\RequirePackage{xspace}
\begin{document}
	\title{Exclusive photoproduction of vector mesons  in proton-lead ultraperipheral collisions at the LHC }
	\author{Ya-Ping Xie}\email{xieyaping@impcas.ac.cn}
	\affiliation{Institute of Modern Physics, Chinese Academy of
		Sciences, Lanzhou 730000, China}
\author{Xurong Chen}\email{xchen@impcas.ac.cn}
\affiliation{Institute of Modern Physics, Chinese Academy of
	Sciences, Lanzhou 730000, China}
	\begin{abstract}
		 Rapidity distributions of vector mesons are computed in dipole model proton-lead
		 ultraperipheral collisions(UPCs)  at the CERN Larger Hadron Collider(LHC). The dipole model framework is implemented in the calculations of cross sections in the photon-hadron interaction.  The bCGC model and Boosted Gaussian wave functions are employed in the scattering amplitude. We obtain predictions of rapidity distributions of $J/\psi$ meson proton-lead ultraperipheral collisions. The predictions give a good descriptions to the experimental data of ALICE. The rapidity distributions of $\phi$, $\omega$ and $\psi(2s)$ mesons in proton-lead ultraperipheral collisions are also presented in this paper.
	\end{abstract}
	\pacs{24.85.+p, 12.38.Bx, 12.39.St, 13.88.+e}
	\maketitle
\section{introduction}
Diffractive production of vector mesons in hadron-hadron and electron-proton collisions can help us study the QCD dynamics and gluon saturation effect at high energy level \cite{Bertulani:2005ru,Baltz:2007kq}. The H1 and ZEUS collaborations have measured the cross sections of $J/\psi$ in diffractive process at HERA \cite{Chekanov:2002xi,Chekanov:2004mw,Aktas:2005xu,Alexa:2013xxa}. The LHCb collaborations have measured the rapidity distributions of $J/\psi$ and $\psi(2s)$ in hadron-hadron ultraperipheral collisions (UPCs) at the LHC\cite{Aaij:2013jxj,Aaij:2014iea,LHCb:2016oce,Abbas:2013oua,Abelev:2012ba,TheALICE:2014dwa,Adam:2015gsa,Adam:2015sia}.  Various theoretical approaches can be found to compute the production of vector mesons in UPCs and diffractive processes \cite{Klein:1999qj,Frankfurt:2002sv,Goncalves:2005yr,Ryskin:1992ui,Toll:2012mb,Adeluyi:2012ph,Lappi:2013am,Xie:2016ino,Xie:2015gdj,Xie:2017mil}.\\
\indent In hadron-hadron UPCs, the direct hadronic interaction is suppressed since the two hadrons barely touch. The photon-induced interaction is dominant in hadron-hadron UPCs. Vector mesons can be produced in photon-induced  process. The dipole model is a phenomenological model in small-x physics \cite{Forshaw:2003ki}, it can describe the photon-hadron interaction very well. In the dipole model, the interaction between virtual photon and hadron can be viewed as three steps.  Firstly, the virtual photon splits into quark and antiquark. Therefore, the quark-antiquark interacts with proton by exchange gluons. Finally, the quark-antiquark recombine into other particles, for example, vector mesons or real photon. The important aspect of dipole model is the cross section of a pair of quark-antiquark scattering off a proton through gluons exchange.  Dipole amplitude is the imaginary part of total $\gamma^* p$ cross section. It is important in the diffractive process to calculate the production of vector mesons since the vector meson can be viewed as a probe of the interaction between the dipole and the proton.  The Golec-Biernat-Wusthoff (GBW) model was firstly proposed to describe the dipole cross section in saturation physics \cite{GolecBiernat:1998js}. The Bartel-Golec-Biermat-Kowalski (BGBK) model is an extensive model of the GBW model considering the gluon density evolution according to DGLAP equation \cite{Bartels:2002cj}. The Color-Glass-Condensate (CGC) model was proposed based on Balitsky-Kovchegov (BK) evolution equation \cite{Iancu:2003ge,Soyez:2007kg,Ahmady:2016ujw}. The bSat and bCGC models are impact parameter dependent dipole models based on the BGBK and CGC models \cite{Kowalski:2003hm,Kowalski:2006hc,Rezaeian:2012ji,Watt:2007nr,Rezaeian:2013tka}. These models contain free parameters which are determined by fit on cross sections of the inclusive production in DIS. \\
 \indent In the photoproduction of vector meson in diffractive process, the light-cone wave functions of photon and vector meson are employed in the amplitude. The light-cone wave function of photon can be computed  analytically, but the light-cone function of the vector meson can't be computed analytically. Phenomenological models are employed for the vector mesons in the calculations. The Boosted Gaussian model is a successful model for $J/\psi$ and excited states.  The production of $J/\psi$ and $\psi(2s)$ can be applied to check the validity of the Boosted Gaussian wave functions.
Using the dipole amplitude and light-cone wave functions of photon and vector meson, the cross section in diffractive process can be easily computed. \\
\indent On other side, the cross sections of heavy vector mesons in diffractive process is investigated in perturbative QCD approach \cite{Jones:2013pga,Jones:2013eda,Jones:2016icr}. The vector meson amplitude is proportional to the gluon density.  The leptonic decay width of the heavy vector meson is included in the amplitude.  The diffractive production also can be computed in momentum space \cite{Ivanov:2004ax,Rybarska:2008pk,Cisek:2010jk,Cisek:2014ala}. In momentum space, the unintegrated gluon distribution is employed in the calculation.\\
\indent In nucleus-nucleus UPCs, the nucleus can remains intact or breaks up. If the nucleus remains intact, the process is coherent process. If the nucleus breaks up, the process is called incoherent process. In previous works \cite{Xie:2017mil}, we have studied the incoherent vector production in lead-lead
 collisions at the LHC. In the previous works, we have presented differential cross sections as a function $|t|$. In the coherent process, the $|t|$ spectra is steep at small $|t|$ region, while the $|t|$ spectra of incoherent production is exponential. In this paper, we focus on vector mesons coherent cross sections of proton-lead UPCs. \\
\indent  In this paper, the  bCGC model is employed to calculate the exclusive photoproduction of vector mesons in photon-lead UPCs.  The aim of this paper is to calculate predictions of exclusive photoproduction of $J/\psi$, $\psi(2s)$, $\omega$ and $\phi$ mesons in proton-lead UPCs at the LHC.    In Section II, the formalism of this work is reviewed. In section III, the numerical results are presented and some discussion are also listed. The conclusions are in section IV.
\section{Vector meson production in the dipole model}
In this paper, we focus on the production of vector meson in proton-lead UPCs. The rapidity distributions of vector
 meson production in UPCs is the product of cross sections of $\gamma+\mathrm{p}\to V+\mathrm{p}$ and  the photon flux factor . The rapidity distributions of vector meson in hadron-hadron UPCs is given as follows\cite{Ducati:2013tva}
\begin{eqnarray}
\frac{d\sigma}{dy}=k^+\frac{dn^{h_2}}{dk^+}(k^+)\sigma^{\gamma h_1\to Vh_1}(W^+)+k^-\frac{dn^{h_1}}{dk^-}(k^-)\sigma^{\gamma h_2\to Vh_2}(W^-).
\label{dsdy}
\end{eqnarray}
In above equation, $k$ is momentum of the radiated photon from hadrons. $\mathrm{y}$ is the rapidity of the vector meson.  $k^{\pm}=M_V/2\exp(\pm |\mathrm{y}|)$.  $\mathrm{W}^{\pm}$ is the center mass energy in diffractive process In UPCs, $W^{\pm}=(2k^\pm\sqrt{s})^{1/2}$ with $\sqrt{s}$ center-energy.  $dn/dk$ is photon flux\cite{Bertulani:2005ru}. Photon flux for proton is given by 
\begin{eqnarray}
\frac{dn}{dk}(k)=\frac{\alpha_{em}}{2\pi k}\Big[1+\Big(1-\frac{2k}{\sqrt{s}}\Big)^2\Big]\Big(
\ln \Omega-\frac{11}{6}+\frac{3}{\Omega}-\frac{3}{2\Omega^2}+\frac{1}{3\Omega^3}\Big),
\end{eqnarray}
where $\Omega=1+0.71\mathrm{GeV}^2/Q^2_{min}$, with $Q^2_{min}=k^2/\gamma^2_L$, $\gamma_L$ is the Lorentz boost factor with $\gamma_L=\sqrt{s}/2m_p$.
The photon flux for the nuclei is given as \cite{Bertulani:2005ru}
\begin{eqnarray}
\frac{dn}{dk}(k) =\frac{2Z^2\alpha_{em}}{\pi k}\Big[\xi K_0(\xi)K_1(\xi)-\frac{\xi^2}{2}(K_1^2(\xi)-K_0^2(\xi)\Big],
\end{eqnarray}
where the $\xi$ is defined as $\xi=k(R_{h_1}+R_{h_2})/\gamma_L$, $K_0(x) $ and $K_1(x)$ are second Bessel functions.\\
\indent The cross sections of $\sigma^{\gamma p\to Vp}(W)$ is integrated by $|t|$ as
\begin{equation}
\sigma^{\gamma p\to Vp}(x_p,Q^2)=\int dt\frac{d\sigma^{\gamma p\to Vp}}{dt}(x_p,Q^2,\Delta).
\end{equation}
Then, the differential cross section of $\gamma+p\to V+p$ is given as \cite{Kowalski:2003hm,Kowalski:2006hc}
\begin{eqnarray}
\frac{d\sigma^{\gamma p\to Vp}}{dt}(x_p,Q^2,\Delta)=\frac{R_g^2(1+\beta^2)}{16\pi^2}
|\mathcal{A}^{\gamma p\to Vp}(x_p,Q^2,\Delta)|^2,
\label{dsigma1}
\end{eqnarray}
with $x_p=\frac{M_V}{\sqrt{s}}\exp(\mp| \mathrm{y|})$ or $x_p=M_V^2/W^2$ and $t=-\Delta^2$. The amplitude $\mathcal{A}^{\gamma p\to Vp}(x_p,Q^2,\Delta)$ in Eq.~(\ref{dsigma1}) is written as
\begin{eqnarray}
\mathcal{A}^{\gamma p\to Vp}(x_p, Q^2,\Delta)= i\int
d^2r\int_0^1\frac{dz}{4\pi} \int
d^2b(\psi_V^*\psi_{\gamma})_{T}(z,r,Q^2)e^{-i(\bm b-(1-z)\bm r)\cdot\bm
	\Delta }\mathcal{N}(x_p,\bm r,\bm b),\notag\\
\label{amp}
\end{eqnarray}
where T denotes the transverse overlap function of photon and vector meson functions with $Q^2=0$, since the photon is real one in UPCs.  And $\beta$ is ratio of the imaginary part amplitude to the real part amplitude. It is written as
\begin{equation}
\beta=\tan (\frac{\pi}{2}\delta),  \quad\text{with}\quad \delta=\frac{\partial \ln (\mathrm{Im}\mathcal{A}(x))}{\partial \ln(1/x)}.
\end{equation}
The factor $R_g^2$ reflects the skewedness \cite{Shuvaev:1999ce}, it gives
\begin{equation}
R_g=\frac{2^{2\delta+3}}{\sqrt{\pi}}\frac{\Gamma(\delta+5/2)}{\Gamma(\delta+4)}.
\end{equation}
The differential cross sections of $\gamma+A\to V+A$ is given as \cite{Lappi:2013am,Xie:2016ino}
\begin{eqnarray}
\frac{d\sigma^{\gamma A\to VA}}{dt}(x_p,Q^2,\Delta)=\frac{R_g^2(1+\beta^2)}{16\pi}
\left|\langle\mathcal{A}^{\gamma A\to VA}(x_p, Q^2,\Delta)\rangle_N\right|^2,\label{dsigma}
\end{eqnarray}
where the average amplitude is calculated as~\cite{Lappi:2013am,Xie:2016ino}
\begin{eqnarray}
\langle\mathcal{A}^{\gamma A\to VA}(x_p, Q^2,\Delta)\rangle_N&=& i\int d^2\bm{r}\int_0^1\frac{dz}{4\pi}
\int d^2\bm{b}(\Psi_V^*\Psi_{\gamma})_{T}(z,r,Q^2)e^{-i(\bm b-(1-z)\bm r)\cdot\Delta}\notag\\
&&\times2(1-\exp(-\frac{\sigma_{q\bar{q}}(x_p,\bm r)}{2}AT_A(\bm b)).
\label{amp}
\end{eqnarray}
Corresponding $T_A(b)$ is the Wood-Saxon distribution, and $\sigma_{q\bar{q}}(x_p,r)$ is integrated impact parameter cross section, it is calculated as 
\begin{eqnarray}
\sigma_{q\bar{q}}(x_p,\bm r)=\int d^2\bm b \mathcal{N}(x_p,\bm r,\bm b),
\end{eqnarray}
where  $\mathcal{N}(x_p,\bm r,\bm b)$ is the dipole amplitude between the dipole and proton.  In the bCGC model, the dipole amplitude is computed as \cite{Iancu:2003ge,Rezaeian:2013tka}
  \begin{eqnarray}\label{dipole}
\mathcal{N}(x,\bm r,\bm b)=2\times\begin{cases}
  \mathcal{N}_0(\frac{rQs}{2})^{2(\gamma_s+(1/\kappa\lambda Y)\ln(2/rQs))},\quad\!\!\! rQs\leqslant 2,\\
  1-\exp\big(-a\ln^2(b rQs)\big),\quad\quad\quad rQs>2,
  \end{cases}
  \end{eqnarray}
  where $Qs(x,\bm b)=(x/x_0)^{\lambda/2}\exp(-\frac{\bm b^2}{4\gamma_sB_p})$, $\kappa=9.9$, and $Y=\ln(1/x)$. $a$ and $b$ are given as
   \begin{equation}
   \begin{split}
   & a=-\frac{\mathcal{N}^2_0\gamma_s^2}{(1-\mathcal{N}_0)^2\ln(1-\mathcal{N}_0)},\\
   &b=\frac{1}{2}(1-\mathcal{N}_0)^{-(1-\mathcal{N}_0)/(2\mathcal{N}_0\gamma_s)}.
   \end{split}
   \end{equation}
   In the bCGC model, $B_p$, $x_0$, $\gamma_s$, $\mathcal{N}_0$ and $\lambda$ are free parameters and they are fitted from the experimental data.
   There are various sets parameters in the literatures. In this paper, we use the same parameters in Ref. \cite{Watt:2007nr,Rezaeian:2013tka}. Parameters of bCGC model used in this paper are presented in TABLE. \ref{IPP2}.  The Fit 3 are parameters with no saturation. It means that the form of Eq. (\ref{dipole}) $rQs\leqslant 2$ is also taken for  $rQs>2$.
     \begin{table}[h]
     	\begin{tabular}{p{1cm}p{2cm}p{2cm}p{2cm}p{2.cm}p{2.5cm}p{3.5cm}p{2.5cm}}
     		\hline
     		\hline
     	&$B_p$/~$\mathrm{GeV}^{-2}$ &$m_{u,d,s}$/GeV&$m_{c}$/GeV& $\gamma_s$ & $N_0$ & $x_0$ & $\lambda$\\
     		\hline
     		Fit 1&$5.5$&$\approx 0$&1.27& 0.6599$\pm$0.0003    & 0.3358$\pm$0.0004&
     		0.00105$\pm$1.13$\times10^{-5}$ &  0.2063$\pm$0.0004    \\
     			Fit 2&$5.5$  &$\approx 0$&1.4    &0.6492$\pm$0.0003 &0.3658$\pm$0.0006  & 
     		0.00069$\pm$6.46$\times10^{-6}$&     0.2023$\pm$0.0003   \\
     		Fit 3&$7.5$  &0.14&1.4    &0.43 &0.565  & 
     		1.34$\times10^{-6}$&     0.109   \\
     		\hline
     		\hline
     	\end{tabular}
     	\caption{Parameters for bCGC model \cite{Watt:2007nr,Rezaeian:2013tka}}
     	\label{IPP2}
     \end{table}
     \\
  \indent The overlap function between photon and vector meson in Eq.~(\ref{amp})  are given as follows
   \begin{eqnarray}
   (\Psi_V^*\Psi_{\gamma})_T(r,z,Q^2)=e_fe\frac{N_c}{\pi z(1-z)}\lbrace  m_f^2
   K_0(\epsilon r)\phi_T(r,z)-(z^2+(1-z)^2)\epsilon K_1(\epsilon r)\partial_r
   \phi_T(r,z)\rbrace,\notag\\
   \end{eqnarray}
   where $e_f$ is effective  charge for quark and $m_f$ is effective quark mass. $\epsilon=\sqrt{z(1-z)Q^2+m_f^2}$ and $\phi_T(r,z)$ is the scalar functions, $K_0(x)$ and $K_1(x)$ are second kind Bessel functions. There is no analytic expression for the scalar functions of the vector mesons. There are some successful models for the scalar functions.
   The Boosted Gaussian model is a phenomenological model. The scalar function of $J/\psi$ in Boosted Gaussian model is written as
   \begin{eqnarray}
   \phi^{1s}_T(r,z)=N_Tz(1-z)\exp\big(-\frac{m_f^2\mathcal{R}^2}{8z(1-z)}-
   \frac{2z(1-z)r^2}{\mathcal{R}^2}+\frac{m_f^2\mathcal{R}^2}{2}\big).
   \end{eqnarray}
  The wave function of $\phi$ and $\omega$ are the same as the function of $J/\psi$ meson.  The scalar function for $\psi(2s)$ meson in the Boosted Gaussian model is given as~\cite{Armesto:2014sma}
   \begin{eqnarray}
   \phi^{2s}_T(r,z)&=&N_Tz(1-z)\exp\big(-\frac{m_f^2\mathcal{R}^2}{8z(1-z)}-
   \frac{2z(1-z)r^2}{\mathcal{R}^2}+\frac{m_f^2\mathcal{R}^2}{2}\big)\notag\\
   &\times&\Big[1+\alpha_{2s}\Big(2+\frac{m_f^2\mathcal{R}^2}{8z(1-z)}-
   \frac{4z(1-z)r^2}{\mathcal{R}^2}-m_f^2\mathcal{R}^2\Big)\Big].
   \end{eqnarray}
There are free parameters of the Boosted Gaussian wave functions, such as $N_T, \mathcal{R}^2 , \alpha_(2s)$. The parameters are determined by the normalization conditions and the lepton decay width. The detail process of the parameters calculations can refer Ref. \cite{Kowalski:2006hc}.  The parameter of $J/\psi $ and $\psi(2s)$ in this paper are taken from the Ref.\cite{Armesto:2014sma}.  The parameters of $\omega$ and $\phi$ ($m_q$=0.01 GeV) are calculated in this paper. They  are presented in Table. \ref{wave}. \\
       \begin{table}[htbp]
       	\begin{tabular}{p{1cm} | p{1.5cm}p{1.5cm}p{1.5cm}p{1.5cm}p{1.5cm}p{1.5cm} p{1.5cm}}
       		\hline
       		\hline
       		meson &$e_f$& mass& $f_V$ & $m_f$& $N_T$& $\mathcal{R}^2$&$\alpha_{2s}$ \\
       		\hline
       		& & GeV & GeV & GeV &   & $\text{GeV}^2 $ & \\
       		\hline
       		$\omega$  &$1/3\sqrt{2}$      & 0.782    & 0.0458  & 0.14      & 0.895     &15.78 & \\
       		$\omega$  &$1/3\sqrt{2}$      & 0.782    & 0.0458  & 0.01     & 1.030     &16.40 & \\
       		$\phi$  &$1/3$      & 1.019    & 0.076  & 0.14    &   0.919    &11.2 & \\
       		$\phi$  &$1/3$      & 1.019    & 0.076  & 0.01    &   1.021    &11.57 & \\
       		$J/\psi$  &$2/3$      & 3.097    & 0.274  & 1.27      & 0.596      &2.45 & \\
       		$J/\psi$  &$2/3$      & 3.097    & 0.274  & 1.40   & 0.57     &2.45 & \\
       		$\psi(2s)$  &$2/3$      & 3.686  & 0.198  & 1.27      & 0.70      &3.72 & -0.61\\
       		$\psi(2s)$  &$2/3$      & 3.686  & 0.198  & 1.40      & 0.67      &3.72 & -0.61\\
       		\hline
       		\hline
       	\end{tabular}
       	\caption{Parameters of the scalar functions of the Boosted Gaussian model for $\omega$, $\phi$, $J/\psi$ and $\psi(2s)$ mesons. Parameters of $\phi$ ( $m_q$=0.14 GeV), $J/\psi$ and $\psi(2s)$ are taken from~\cite{Kowalski:2006hc,Armesto:2014sma}.}
       	\label{wave}
       \end{table}
\indent Using the parameters of the bCGC model and the Boosted Gaussian wave functions. The rapidity distributions of the vector mesons in proton-lead UPCs can be calculated according to Eq. (\ref{dsdy}). In the Section III, the predictions of rapidity distributions will be computed for four kinds vector mesons. 
 \section{Numerical results and discussions}
\begin{figure}[!h]
  	\centering
  	\includegraphics[width=3in]{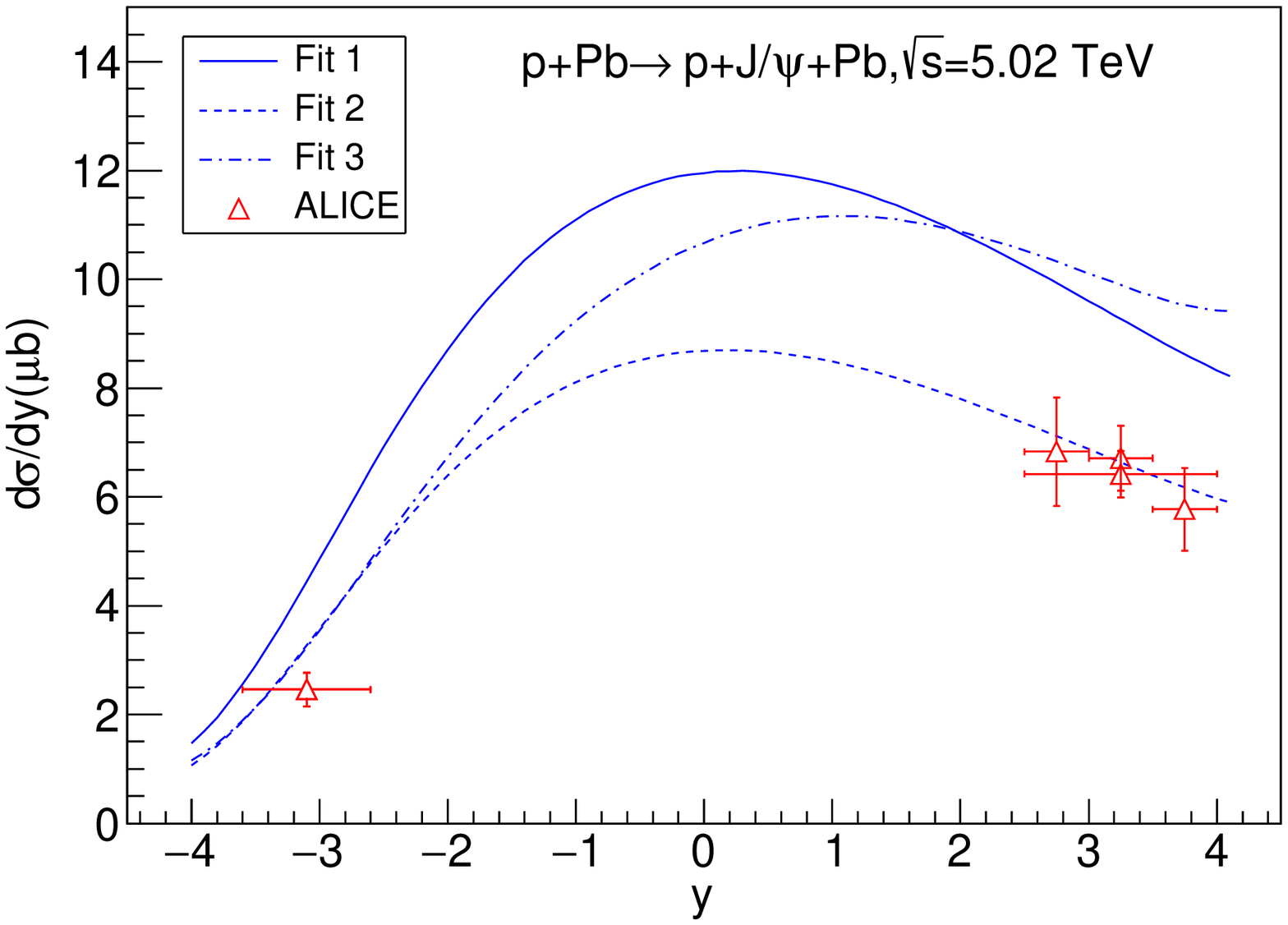}
  	\includegraphics[width=3in]{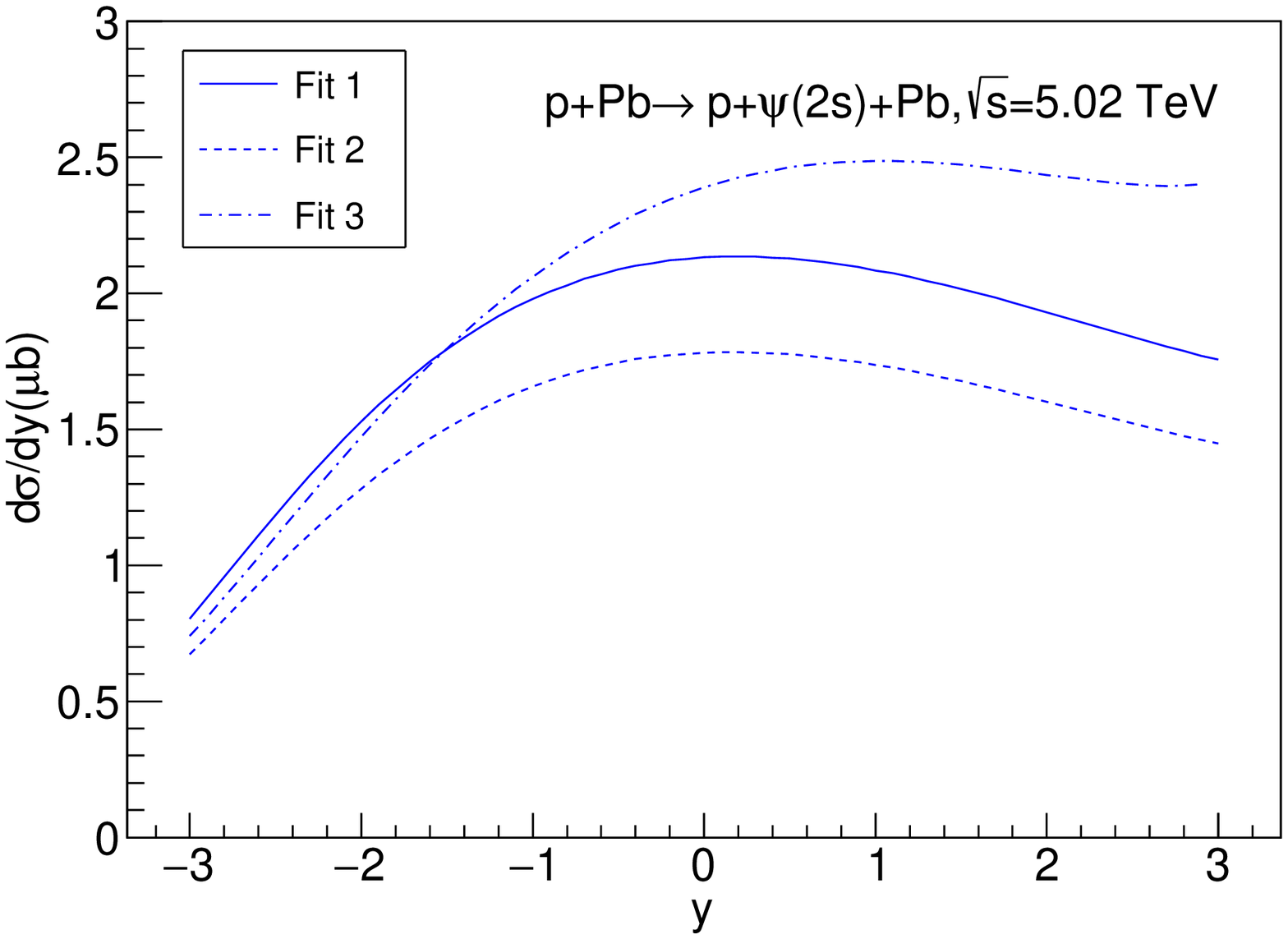}
  	\caption{(Color online) Predictions of rapidity distributions of $J/\psi$ and $\psi(2s)$ mesons in proton-lead ultraperipheral collisions at the LHC compared with the experimental data of the ALICE collaboration\cite{TheALICE:2014dwa}. The solid lines  are predictions using parameters  Fit 1 and the dashed lines are predictions using parameters Fit 2, the dotted-dashed lines are predictions using parameter Fit 3 (no saturation). }
  	\label{chams}
  \end{figure}

In this section, the prediction of rapidity distributions of vector mesons in proton-lead UPCs will be presented. We compute the rapidity distributions of $J/\psi$ and $\psi(2s)$ mesons in proton-lead UPCs. The parameters in bCGC model are taken from \cite{Rezaeian:2013tka} in the calculations. The parameters of the Boosted Gaussian wave functions of $J/\psi$ and $\psi(2s)$ are used as the same as Ref \cite{Armesto:2014sma}.  The rapidity distributions of $J/\psi$ is shown in left graph in Fig.~\ref{chams}.  The solid lines are predictions using the parameters of bCGC model with $m_c=1.27 $ GeV in Ref \cite{Rezaeian:2013tka} and the dashed lines are predictions using the parameters of bCGC model with $m_c$=1.4 GeV. The dotted-dashed lines are prediction using no saturation model. It can be seen that the no saturation prediction is larger than the saturation model with same wave functions ($m_c$=1.4 GeV) . The  experimental data of $J/\psi$ in proton-lead UPCs are also presented in the figure. It can be seen that the theoretical prediction in this work with saturation model give a good description to the experimental data. But the no saturation model overshoots the experimental data.  Thus, we can extend the same method to calculate the other vector mesons rapidity distributions in proton-lead UPCs. The results of $\psi(2s)$ is shown in right graph of Fig.~\ref{chams}. Predictions of $\psi(2s)$ can be referred for the future experimental at the LHC.\\
 \begin{figure}[!h]
 	\centering
 	\includegraphics[width=3in]{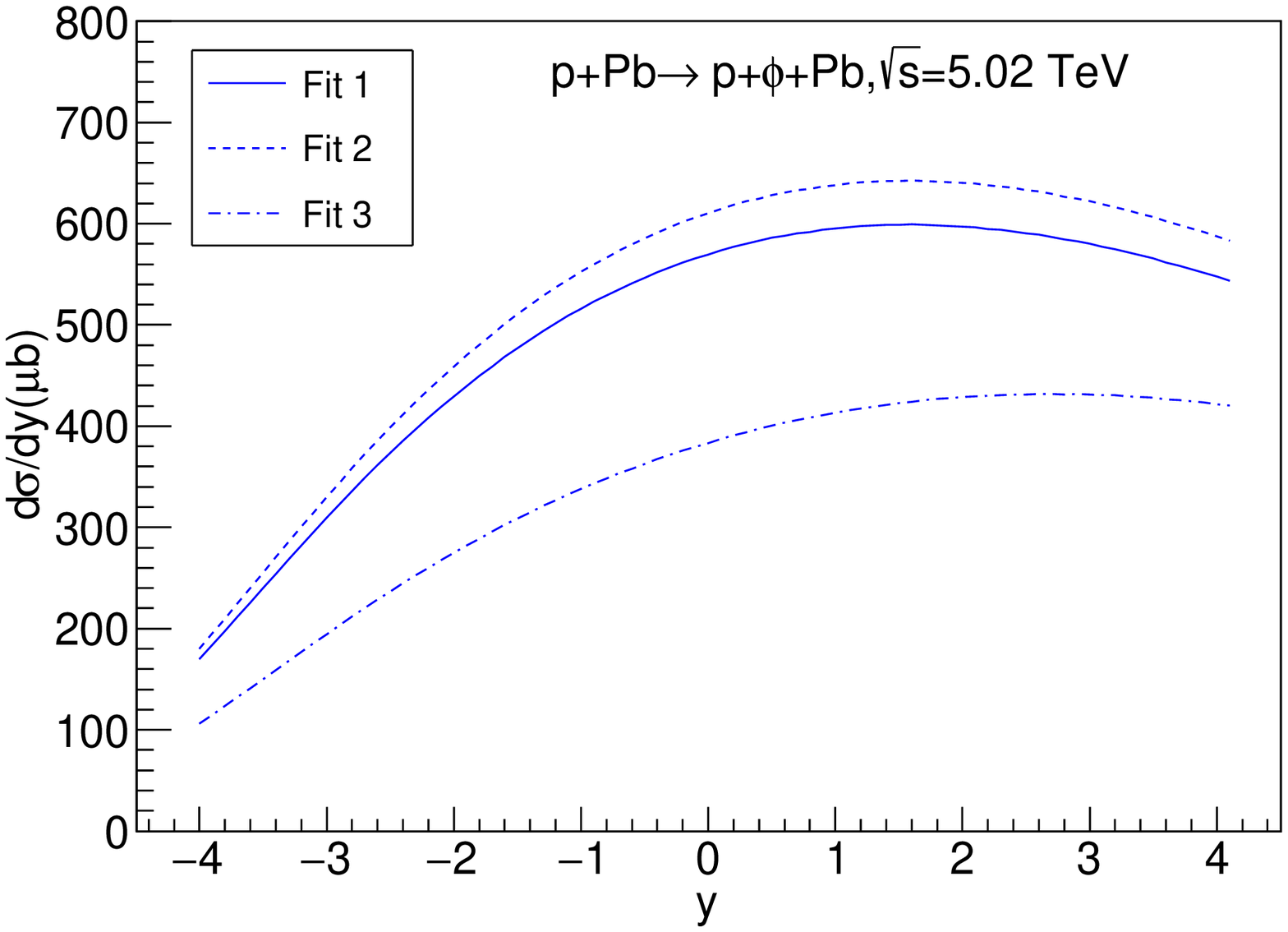}
 	\includegraphics[width=3in]{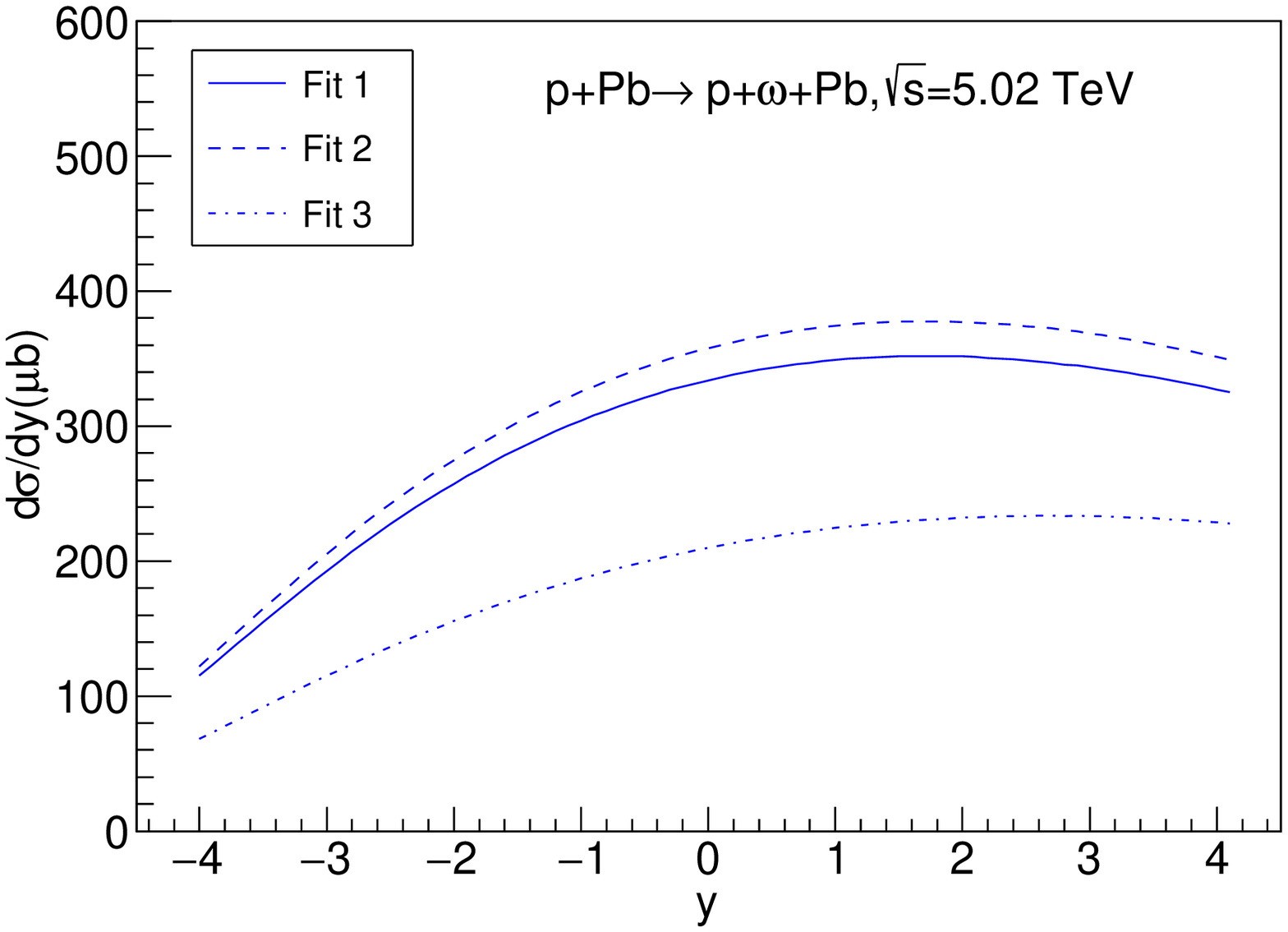}
 	\caption{(Color online) Predictions of rapidity distributions of $\phi$  meson computed in bCGC model using the Boosted Gaussian wave function in proton-proton and proton-lead ultraperipheral collisions at the LHC. The solid lines are predictions using parameters  Fit 1 and the dashed lines are predictions using parameters Fit 2, the dotted-dashed lines are predictions using parameter Fit 3 (no saturation). }
 	\label{phi}
 \end{figure}
\indent Besides the heavy vector meson, the rapidity distributions of $\phi$ and $\omega$ mesons in proton-lead UPCs are computed in the bCGC model with Boosted Gaussian wave function in this paper. The predictions are shown in Fig.~\ref{phi} where the solid lines are results using parameters of Fit 1, the dashed lines are rapidity distributions using parameters of Fit 2 and dotted-dashed lines are predictions using parameter FIt 3 (no saturation). It can be seen that the no saturation prediction are smaller than the saturation model since the wave function parameters are different. In the calculations of Fit 1 and 2, the wave function parameters are using $m_q$=0.01  GeV and in the calculations of Fit 3, the wave function parameters are using $m_q$=0.14 GeV.  We hope that the experimental data will be measured in the future. We can compare the theoretical prediction with the experimental data.\\
  \begin{figure}[!h]
  	\centering
  	  	\includegraphics[width=3.5in]{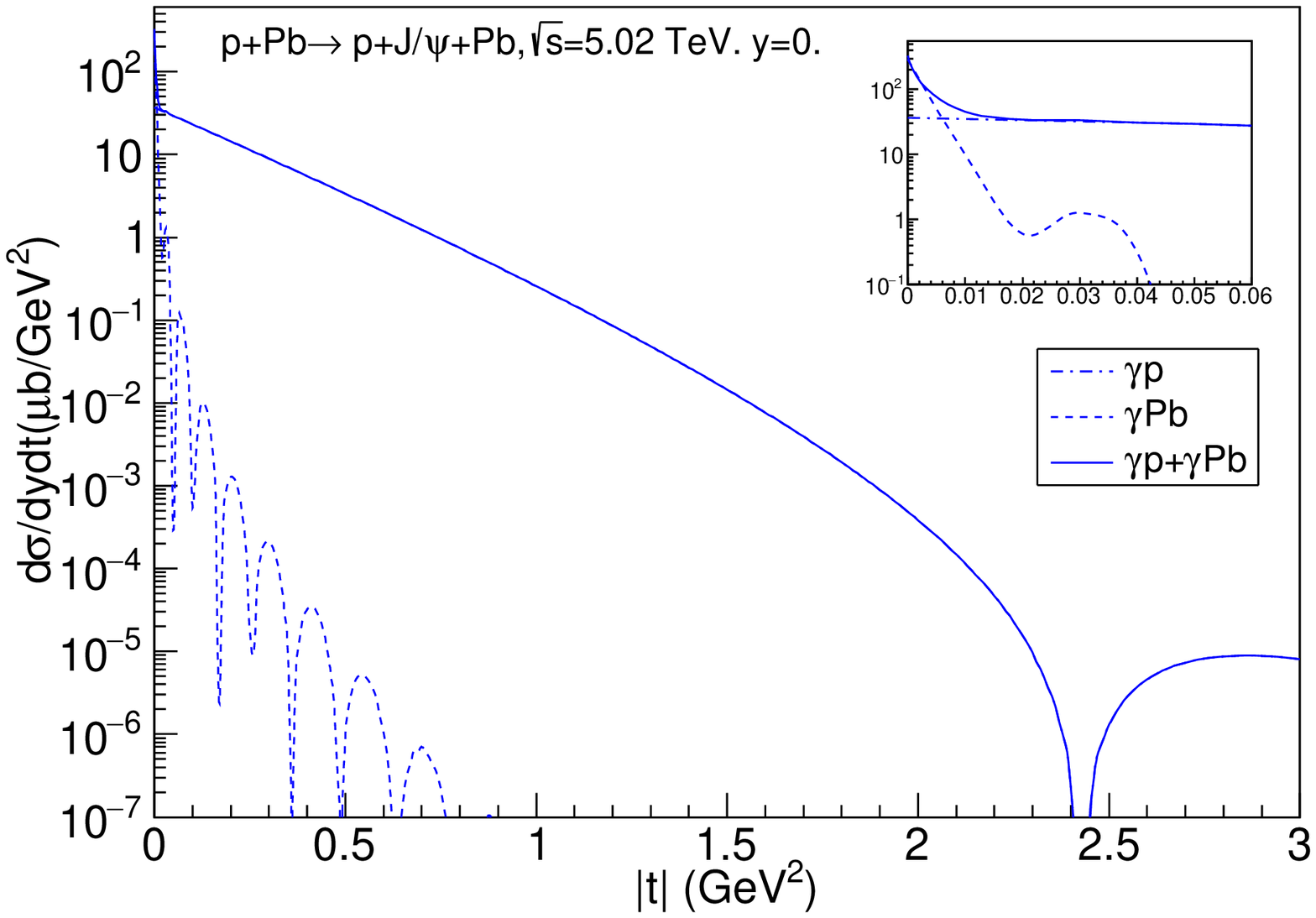}
  	\includegraphics[width=3.5in]{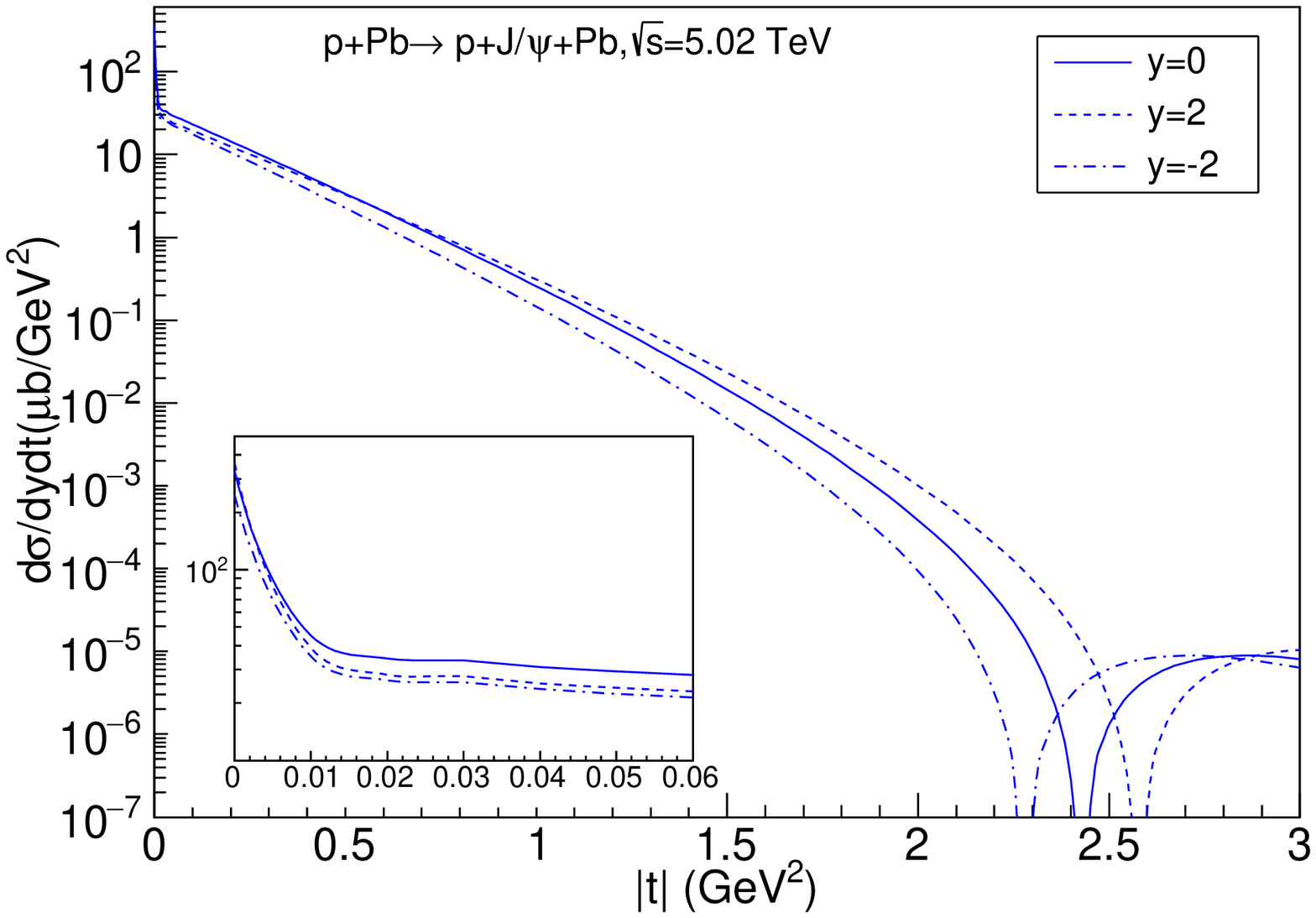}
  	\caption{(Color online) Predictions of $J/\psi$ differential cross section as a function of $|t|$ at fixed rapidities in proton-lead UPCs. The
  		bCGC parameters are taken with Fit 2. The small boxes are regional enlarged figures.}
  	\label{dsdt}
  \end{figure}
  \indent At the end of the day, we present the differential cross sections of $J/\psi$ in proton-lead UPCs. The numerical results are shown in FIg. \ref{dsdt}.  The left graph is differential cross sections at y=0. The dashed line is the results of $d\sigma^{\gamma+Pb\to J/\psi+Pb }/dydt$ and dotted-dashed line is results of $d\sigma^{\gamma+p\to J/\psi+p}/dydt$. The solid line is the sum of two channel cross sections. It can be seen that the cross section of $\gamma+Pb$ can be neglected expect the small $|t|$.  In the $|t|=0\;\mathrm {GeV}^2$, there is a peak as descried from the left graph. The right graph are total differential cross sections at the different rapidities. The solid line is results at y=0; the dashed line describes differential cross sections at y=2; respectively, the dotted-dashed line is predictions at y=-2. We can see that the differential cross sections at y=2 are larger than the results at y=-2. This asymmetry is reflected in the rapidity distributions. In the lead-lead UPCs, the coherent cross sections are symmetry as described in our previous work\cite{Xie:2017mil}.  We hope the experiment can measure the $|t|$ distributions in UPCs and the theoretical prediction can be compared with experimental data.  In Ref \cite{Adamczyk:2017vfu,Klein:2017vua} STAR collaboration have measured differential cross sections of $\rho$ meson in UPCs. We hope the LHC also can measure the $|t|$ spectrum of $J/\psi$ in the future. 
\section{conclusion}
In this paper, we have studied the exclusive photoproduction of $J/\psi$, $\psi (2s)$, $\phi$ and $\omega$  in proton-lead UPCs at the LHC.  The bCGC model and the Boosted Gaussian wave function are employed in the calculation. The theoretical predictions of $J/\psi$  meson rapidity distributions are compared with the experimental data measured by the ALICE collaboration.  It can be seen that the predictions of $J/\psi$ using saturation model give a good description to the experimental data. Predictions of rapidity distribution of $\psi(2s)$,$\phi$ and $\omega$ are also computed in proton-lead UPCs at the LHC  in this paper. Since there is no experimental data for the rapidity distributions for the other three vector mesons now. We hope these theoretical predictions can be compared in the future.  The differential cross sections of vector mesons as a function $|t|$ are also calculated in this paper. From the differential cross sections, we can see that the $\gamma Pb$ channel contributes a peak when $|t|=0\; \mathrm{GeV}^2$  and It can be neglected when the $|t|$ is larger  0.02 $\mathrm{GeV}^2$. We hope the $|t|$ spectra can be also checked by the experiments at the LHC. 
\section{Acknowledgements}
This work is supported in part by the National 973 project in China (No:~2014CB845406).

\end{document}